\documentclass[smallextended]{svjour3} 
\usepackage{amsmath,amssymb}
\usepackage{graphics,epsfig,psfrag}
\usepackage{float}

\begin{document}

\title{Application of the Gauss-Bonnet theorem to lensing in the NUT
metric}

\author{Mourad Halla         \and
        Volker Perlick
}

\institute{M. Halla \at
              ZARM, University of Bremen \\
              28359 Bremen, Germany \\
              \email{mourad.halla@zarm.uni-bremen.de} 
           \and
           V. Perlick \at
              ZARM, University of Bremen \\
              28359 Bremen, Germany \\
              \email{perlick@zarm.uni-bremen.de}
}

\date{Received: date / Accepted: date}

\maketitle

\begin{abstract} 
We show with the help of Fermat's principle that every lightlike geodesic
in the NUT metric projects to a geodesic of a two-dimensional
Riemannian metric which we call the optical metric. The optical
metric is defined on a (coordinate) cone whose opening angle
is determined by the impact parameter of the lightlike geodesic.
We show that, surprisingly, the optical metrics on
cones with different opening angles are locally (but not globally)
isometric. With the help of the Gauss-Bonnet theorem we 
demonstrate that the deflection angle of a lightlike geodesic
is determined by an area integral over the Gaussian curvature
of the optical metric. A similar result is known to be true for 
static and spherically symmetric spacetimes. The generalisation 
to the NUT spacetime, which is neither static nor spherically
symmetric (at least not in the usual sense), is rather non-trivial.             
\end{abstract}

\section{Introduction}\label{sec:intro}
The NUT metric is a solution to the vacuum Einstein equation that 
was found by Newman, Unti and Tamburino (NUT) in 1963
\cite{NewmanTamburinoUnti1963}. It describes the vacuum 
spacetime around a source that is characterised by two parameters, 
called $m$ and $l$ in the following. $m$ detemines the mass of the 
central object, whereas 
$l$ determines its gravitomagnetic charge. In an analogy to
electromagnetism, $m$ corresponds to the electric charge whereas
$l$, which is also known as the NUT parameter,  corresponds to a 
magnetic (monopole) charge. 

The NUT metric has so many unusual properties that 
Misner \cite{Misner1967} called it a ``counter-example to almost anything''.
This, however, does not necessarily mean that the NUT metric is unphysical.
On the contrary, history has taught us to take solutions to Einstein's
vacuum equation seriously, even if they have apparently ``exotic''
properties. So we believe that the existence in Nature
of sources with a non-zero NUT parameter is a possibility that
should not be rejected off-hand.     

Following this line of thought, we have to ask ourselves what kind
of observable features a hypothetical NUT source would have. In 
essence, there are two types of such features. Firstly, we could consider 
the influence of such a source on the motion of massive particles that 
come close to it, secondly we could consider its influence on light rays.
For the discussion of observable effects of a NUT source on massive
particles we refer in particular to papers by Hackmann and 
L{\"a}mmerzahl \cite{HackmannLaemmerzahl2012} and Jefremov and 
Perlick \cite{JefremovPerlick2016}.
We will not discuss this subject here. As to the influence on light rays,
i.e., to the lensing features of a NUT source, there is a paper by
Nouri-Zonoz and Lynden Bell \cite{NouriZonoz:1998va}. It is
the purpose of the present paper to develop their analysis somewhat
further.

Our main goal is to find out whether the lensing features
in the NUT spacetime can be characterised in terms of two-dimensional
Riemannian geometry. It is well known that such a characterisation is
possible in spacetimes that are static and spherically symmetric: In this
case it suffices to consider lightlike geodesics in the equatorial plane $\vartheta
= \pi/2$ because all the other ones are then determined by the spherical
symmetry. Here $(t, r , \vartheta , \varphi )$ are the spherical polar
coordinates in which a static and spherically symmetric spacetime is
usually given. Then one finds that the spatial paths, in the two-dimensional
manifold with coordinates $(r, \varphi )$,  of lightlike geodesics are
the geodesics of a Riemannian metric which is known as the \emph{Fermat
metric} or the \emph{optical metric}. Gibbons and Werner \cite{Gibbons:2008rj}
have used the Gauss-Bonnet theorem to show how the optical metric 
determines lensing features. In particular, they have demonstrated that the
deflection angle is determined by the Gaussian curvature of the optical
metric. In this paper we want to investigate if and how this result
carries over to the NUT spacetime. As the NUT spacetime is not static but
only stationary, and as in the NUT spacetime geodesics are
not contained in a plane, this makes several non-trivial modifications necessary. 

The paper is organised as follows. In Sec.~\ref{sec:NUT} and Sec.~\ref{sec:geo}
we summarise some basic features of the NUT spacetime and its geodesics.
Throughout we do this for a version of the NUT metric that is more general
than the one introduced in the original NUT paper \cite{NewmanTamburinoUnti1963}
because it depends not only on the above-mentioned parameters $m$ and $l$
but also on a third parameter, $C$, that was introduced in 2005 by Manko and
Ruiz \cite{MankoRuiz2005}. In Sec.~\ref{sec:cone} and Sec.~\ref{sec:defl}
we make use of the fact that in the NUT spacetime every lightlike geodesic
is contained in a cone whose opening angle is determined by the impact parameter 
of the lightlike geodesic and we calculate the deflection angle. For the original
NUT metric with $C=-1$ the fact that every geodesic is contained in a cone was
derived already by Zimmerman and Shahir \cite{ZimmermanShahir1989}; this
fact plays a crucial role also in the above-mentioned paper by Nouri-Zonoz and
Lynden-Bell \cite{NouriZonoz:1998va} who consider the NUT metric with $C=0$.
In Sec.~\ref{sec:optical} we use Fermat's principle to define a Riemannian
metric, which we call the optical metric,  on each of these cones and we demonstrate 
that a lightlike geodesic with the corresponding impact parameter projects, 
indeed, to a geodesic of the optical metric. We illustrate the optical metric by 
way of an embedding diagram, we demonstrate that it has negative Gaussian
curvature and we show that the optical metrics on cones with different opening
angles are locally isometric. Finally, in Sec.~\ref{sec:GB} we use the 
Gauss-Bonnet theorem to rewrite 
the deflection angle as an integral over the Gaussian curvature of the optical
metric. As the latter is negative, this result demonstrates that all light rays are
deflected towards the centre and that the deflection angle increases with
decreasing impact parameter.
  
We use the metric signature (-,+,+,+) and we set the speed of light $c= 1$.
We use Einstein's summation convention with greek indices for the four 
spacetime coordinates and with latin indices for the two coordinates of the
two-dimensional manifold on which the optical metric lives.

\section{The NUT metric}\label{sec:NUT}
The NUT metric  \cite{NewmanTamburinoUnti1963} is a 
solution to Einstein's vacuum field equation. It depends on two
parameters both of which have the dimension of a length, a mass
parameter $m$ and a gravitomagnetic charge, also known as the 
NUT parameter, $l$. We assume $m>0$ and $- \infty < l < \infty$
throughout. For $l=0$ the NUT metric reduces to the Schwarzschild
metric.  In 2005 Manko and Ruiz \cite{MankoRuiz2005} brought
forward a generalised version of the NUT metric which involves a 
third parameter, $C$, that is dimensionless and may take any value 
$- \infty < C < \infty$. With the Manko-Ruiz parameter included, 
the metric reads     
\[
g_{\mu \nu} dx^{\mu} dx ^{\nu} = 
-\frac{( r^2- 2 m r-l^2)}{(r^2+ l^2)} 
\Big( dt - 2 l (\mathrm{cos} \, \vartheta + C ) d \varphi \Big) ^2 
\]
\begin{equation}
+ \dfrac{(r^2+l^2) \, dr^2}{r^2-2 m r -l^2}
+ (r^2+ l^2) \Big( d\vartheta^2 + \mathrm{sin} ^2 \vartheta \, d \varphi ^2 \Big)
\, .
\label{metric}
\end{equation}
Here, $(t, r, \vartheta , \varphi )$ are Boyer-Lindquist-type coordinates. We 
assume that $t$ ranges over all of $\mathbb{R}$, $\vartheta$ and $\varphi$ 
are standard coordinates on the two-sphere, and the radius coordinate is
restricted to the domain $m+\sqrt{m^2+l^2} < r < \infty$. At 
$r=m+\sqrt{m^2+l^2}$ the metric features a black-hole horizon. If 
analytically extended beyond the horizon, in the domain $m - \sqrt{m^2+l^2}
< r < m + \sqrt{m^2+l^2}$ the metric is isometric to a cosmological 
solution that was found already in 1951 by Taub \cite{Taub1951}. The 
entire spacetime is therefore known as the Taub-NUT spacetime. In this 
paper, however, we will not consider the Taub region.

By a coordinate transformation $t'=t-2 l C \varphi$, $r'=r$, $\varphi ' = \varphi$,
$\vartheta ' = \vartheta$ the metric (\ref{metric}) is transformed to the 
corresponding metric with $C=0$. This, however, is not a globally well-defined
transformation: As we assume that $\vartheta$ and $\varphi$ are standard
coordinates on the sphere, $\varphi$ is $2 \pi$-periodic; therefore, the transformation
is well-defined only locally, on any neighbourhood that does not include an
entire $\varphi$-line, unless we make the time coordinate periodic with
period $4 \pi | l C|$. As the latter would lead to the most drastic causality 
violation, giving a closed timelike curve through each point of the considered
domain of the spacetime, we will not do this. Therefore, NUT spacetimes with
different values of $C$ are locally but not globally isometric. 

The original NUT metric \cite{NewmanTamburinoUnti1963} is the
metric (\ref{metric}) with $C= -1$. It features a conic singularity on the
half-axis $\vartheta = \pi$ which was interpreted by Bonnor \cite{Bonnor1969}
as a spinning rod. For $C=1$ the singularity is on the other half-axis 
$\vartheta = 0$, and for all other values of $C$ it is on both half-axes,
symmetrically distributed for $C=0$ and asymmetrically for $C \neq 0$. 
For more details on the physical and geometrical meaning of the parameter
$C$ we refer to Manko and Ruiz \cite{MankoRuiz2005} and also to
Jefremov and Perlick \cite{JefremovPerlick2016}.

For any values of $m$, $l$ and $C$ the NUT metric admits a four-dimensional
Lie algebra of Killing vector fields, i.e., a NUT metric with $l \neq 0$ has as many
symmetries as the Schwarzschild metric. For $C= -1$ the Killing vector fields have 
been given already in the original NUT article \cite{NewmanTamburinoUnti1963}
and the resulting symmetry properties have been discussed in detail by
Misner \cite{Misner1963} who advocated the idea of making the $t$ coordinate
periodic. Here we give the four linearly independent Killing vector fields for 
the NUT metric with an arbitrary value of $C$:
\begin{equation}
\xi _0 = \partial _t \, ,
\label{eq:Killing0}
\end{equation} 
\begin{equation}
\xi _1 = - \mathrm{sin} \, \varphi \, \partial _{\vartheta}
- \dfrac{\mathrm{cos} \, \varphi}{\mathrm{sin} \, \vartheta} \Big(
\mathrm{cos} \, \vartheta \, \partial _{\varphi}
+
2l \big( 1 + C \, \mathrm{cos} \, \vartheta \big) \partial _t \Big)
\, ,
\label{eq:Killing1}
\end{equation} 
\begin{equation}
\xi _2  =  \mathrm{cos} \, \varphi \, \partial _{\vartheta}
- \dfrac{\mathrm{sin} \, \varphi}{\mathrm{sin} \, \vartheta} \Big(
\mathrm{cos} \, \vartheta \, \partial _{\varphi}
+
2l \big( 1 + C \, \mathrm{cos} \, \vartheta \big) \partial _t \Big)
\, ,
\label{eq:Killing2}
\end{equation} 
\begin{equation}
\xi _3 = \partial _{\varphi} + 2 l C \partial _t
\, .
\label{eq:Killing3}
\end{equation} 
The Lie brackets are
\begin{equation}
\big[ \xi _0 , \xi _1 \big] = 
\big[ \xi _0 , \xi _2 \big] = 
\big[ \xi _0 , \xi _3 \big] = 0 \, ,
\label{eq:Lie1}
\end{equation}
\begin{equation}
\big[ \xi _1 , \xi _2 \big] = - \xi _3 \, , \:
\big[ \xi _2 , \xi _3 \big] = - \xi _ 1 \, , \:
\big[ \xi _3 , \xi _1 \big] = - \xi _2  \, .
\label{eq:Lie2}
\end{equation}
This demonstrates that $\xi _1$, $\xi _2$ and $\xi _3$  generate a 
three-dimensional group of isometries which is isomorphic to the
rotation group $SO(3,\mathbb{R})$, whereas $\xi _0$ generates
a one-dimensional group of isometries that expresses stationarity.
Note that for $l \neq 0$ the orbits of the rotations are \emph{not} 
two-dimensional spacelike spheres; they are rather three-dimensional
submanifolds with topology $\mathbb{R} \times S^2$ and the 
metric has the signature $(-++)$ on these submanifolds.  For this
reason many authors find it inappropriate to call the NUT metric
``spherically symmetric''. It is, however, safe to call it ``rotationally
symmetric about any radial direction'' or to say that ``it admits an 
$SO(3 , \mathbb{R} )$ symmetry''. 
 
\section{Geodesics in the NUT metric}\label{sec:geo}
With the Killing vector fields known, it is easy to solve the geodesic
equation in the NUT spacetime. For any geodesic $x^{\mu} (s)$,
parametrised by an affine parameter $s$, each of the four Killing
vector fields $\xi _A$ gives us a constant of motion $g_{\mu \nu}
\xi _A ^{\mu} \dot{x}{}^{\nu}$:
\begin{equation}
E = -g_{\mu \nu} \xi_0 ^{\mu} \dot{x}{}^{\nu} =
\dfrac{r^2-2mr-l^2}{r^2+l^2} \, 
\Big( 
\dot{t} - 2l \big( \mathrm{cos} \, \vartheta +C \big) \, \dot{\varphi} 
\Big)    
\, ,
\label{eq:E}
\end{equation}

\[
J_1 = g_{\mu \nu} \xi_1 ^{\mu} \dot{x}{}^{\nu} =
- \mathrm{sin} \, \varphi \, (r^2+l^2) \, \dot{\vartheta} 
\]
\begin{equation}
-
\mathrm{cos} \, \varphi \, \mathrm{cos} \, \vartheta \,
\mathrm{sin} \, \vartheta \, (r^2+l^2) \, \dot{\varphi}
+2 \, l \, E \, \mathrm{cos} \, \varphi \, \mathrm{sin} \, \vartheta
\, ,
\label{eq:J1}
\end{equation}

\[
J_2 = g_{\mu \nu} \xi_2 ^{\mu} \dot{x}{}^{\nu} =
\mathrm{cos} \, \varphi \, (r^2+l^2) \, \dot{\vartheta} 
\]
\begin{equation}
-
\mathrm{sin} \, \varphi \, \mathrm{cos} \, \vartheta \,
\mathrm{sin} \, \vartheta \, (r^2+l^2) \, \dot{\varphi}
+2 \, l \, E \, \mathrm{sin} \, \varphi \, \mathrm{sin} \, \vartheta
\, ,
\label{eq:J2}
\end{equation}

\begin{equation}
J_3 = g_{\mu \nu} \xi_3 ^{\mu} \dot{x}{}^{\nu} =
\mathrm{sin}{}^2 \vartheta \, (r^2+l^2) \, \dot{\varphi}
+2 \, l \, E \, \mathrm{cos} \, \vartheta
\, .
\label{eq:J3}
\end{equation}
Hence
\begin{equation}
J^2 := J_1^2 +J_2^2+J_3^2 = 
 (r^2+l^2) \, \big( \dot{\vartheta}{}^2 + \mathrm{sin}{}^2 \vartheta \,
 \dot{\varphi}{}^2 \big) + 4 \, l^2 E^2
\label{eq:J}
\end{equation}
and
\begin{equation}
\begin{pmatrix}
\, J_1 \, \\ J_2 \\ J_3
\end{pmatrix}
\boldsymbol{\cdot}
\begin{pmatrix}
\, \mathrm{cos} \, \varphi \, \mathrm{sin} \, \vartheta \, 
\\ 
\, \mathrm{sin} \, \varphi \, \mathrm{sin} \, \vartheta \, \\ 
\mathrm{cos} \, \vartheta 
\end{pmatrix}
=
2 \, l \, E 
\, .
\label{eq:cone}
\end{equation}
The latter equation, in which the central dot means the standard
Euclidean scalar product in $\mathbb{R}{}^3$, demonstrates that
a geodesic with constants of motion $E$, $J_1$, $J_2$ and 
$J_3$ is contained in a cone
whose vertex is at the coordinate origin, whose symmetry axis is
spanned by the vector with Cartesian components
$J_1$, $J_2$ and $J_3$ and whose opening angle $\alpha$ 
is given by 

\begin{equation}
\mathrm{cos} \, \alpha =  2 \, l \, E/J
\, .
\label{eq:angle}
\end{equation}
\\
Note that this statement refers to a three-dimensional Euclidean space
that is defined by the chosen coordinates. We will later discuss the 
intrinsic geometries of these coordinate cones.

For the NUT metric with $C= -1$, the fact that each geodesic is 
contained in a cone was discovered already by Zimmerman and 
Shahir \cite{ZimmermanShahir1989}. (Note, however, that in
their paper the $g_{tt}$ component of the metric is misprinted.)
Eq. (\ref{eq:angle}) demonstrates that the opening angle of this
cone is independent of $C$ if the geodesic is labelled by the 
constant of motion $J/E$. Of course, for $l=0$ (\ref{eq:angle})
gives $\alpha = \pi /2$, i.e., it reproduces the well-known fact
that in the Schwarzschild spacetime each geodesic is contained
in a plane through the coordinate origin. 

In addition to the constants of motion that arise from the Killing
symmetries, also the Lagrangian
\begin{equation}
\mathcal{L} = 
\dfrac{1}{2} \, g_{\mu \nu} \dot{x}{}^{\mu} \, \dot{x}{}^{\nu}
\label{eq:L}
\end{equation}
is a constant of motion. With the help of the four constants of 
motion $E$, $J^2$, $J_3$ and $\mathcal{L}$ the geodesic 
equation can be written in first-order form. To that end we have
to solve, in this order, (\ref{eq:J3}), (\ref{eq:E}), (\ref{eq:J})
and (\ref{eq:L}) for $\dot{\varphi}$, $\dot{t}$, $\dot{\vartheta}{}^2$
and $\dot{r}{}^2$ which results in
\begin{equation}
\dot{\varphi} = 
\dfrac{
J_3-2 \, l \, E\, \mathrm{cos} \, \vartheta
}{
(r^2+l^2) \, \mathrm{sin} ^2 \vartheta
}
\, ,
\label{eq:geo1}
\end{equation}
       
\begin{equation}
\dot{t} = 
\dfrac{(r^2-2mr-l^2) \, E}{r^2+l^2}
+ 2 \, l \, \big( \mathrm{cos} \, \vartheta + C \big)
\dfrac{
\big( J_3-2 \, l \, E \, \mathrm{cos} \, \vartheta \big)
}{
(r^2+l^2) \, \mathrm{sin} ^2 \vartheta
}
\, ,
\label{eq:geo2}
\end{equation}

\begin{equation}
\dot{\vartheta}{}^2 = 
\dfrac{
J^2 - 4 \, l^2 E^2}{(r^2+l^2)^2}
-
\dfrac{ 
( J_3-2 \, l \, E\, \mathrm{cos} \, \vartheta )^2
}{
(r^2+l^2) \, \mathrm{sin} ^2 \vartheta
}
\, ,
\label{eq:geo3}
\end{equation}

\begin{equation}
\dot{r}{}^2 = 
E^2 - \dfrac{(r^2-2mr-l^2)}{(r^2+l^2)} \, 
\big( J^2- 4 \, l^2 E^2 \big) 
+ \dfrac{2 \, \mathcal{L} \big( r^2 - 2 m r  l^2 \big)}{(r^2+l^2)}
\, .
\label{eq:geo4}
\end{equation}
Here $J^2-4 \, l^2E^2$ is the \emph{Carter constant} 
which arises as a separation constant if the geodesic equation
is written in Hamilton-Jacobi form. Note that the spatial paths
of the geodesics, which are determined by eqs. (\ref{eq:geo1}),
(\ref{eq:geo3}) and (\ref{eq:geo4}), are not affected by the
Manko-Ruiz parameter $C$.

\section{Lightlike geodesics contained in a cone with 
$\vartheta = \mathrm{constans}$}\label{sec:cone}
In the Schwarzschild spacetime it suffices to consider geodesics with 
$J_1=J_2=0$, which are exactly the geodesics in the equatorial plane 
$\vartheta = \pi /2$. If one has determined all such geodesics, one gets 
all the other ones by applying all possible rotations, i.e., the action of 
the group generated by the Killing vector fields $\xi _1$, 
$\xi _2$ and $\xi _3$. An analogous statement holds for the NUT metric:
Also in this case it suffices to consider geodesics with $J_1 = J_2 =0$ which 
now are the geodesics contained in cones of the form $\vartheta = 
\mathrm{constans}$; the opening angle $\alpha$ of such a cone equals 
the coordinate angle $\vartheta$. Again, one gets all the other geodesics 
by applying to such geodesics all possible rotations. 

If $J_1 = J_2 =0$ and $\vartheta = \mathrm{constans}$, (\ref{eq:J1}) 
and (\ref{eq:J2}) require
\begin{equation}
\dot{\varphi} = \dfrac{2 \, l \, E}{(r^2+l^2) \, \mathrm{cos} \, \vartheta}
\label{eq:dphitheta}
\end{equation}
\\
and (\ref{eq:cone}) yields
\begin{equation}  
\dfrac{J_3}{E}  = \dfrac{2 \, l}{\mathrm{cos} \, \vartheta} .
\label{eq:J3theta}
\end{equation}
The latter equation demonstrates that, in a NUT spacetime with $l \neq 0$,
the opening angle $\vartheta$ of the cone is determined by the 
\emph{impact parameter} $J_3/E$ and vice versa. In the Schwarzschild case
$l=0$ eqs. (\ref{eq:dphitheta}) and (\ref{eq:J3theta}) determine
neither $\dot{\varphi}$ nor $J_3/E$ because with $l$ also 
$\mathrm{cos} \, \vartheta$ goes to zero.

For lightlike geodesics ($\mathcal{L} = 0$), inserting (\ref{eq:J3theta})
into (\ref{eq:geo4}) results in
\begin{equation}
\dot{r}{}^2 = E^2 \Bigg( 1 - \dfrac{4 \, l^2 (r^2-2mr-l^2)}{(r^2+l^2)^2}
\, \mathrm{tan}{}^2 \vartheta \Bigg) \, .
\label{eq:drtheta}
\end{equation}
Differentiating this equation with respect to the affine parameter and dividing
by $2 \dot{r}$ gives
\begin{equation}
\ddot{r} = - \dfrac{4 \, l^2 E^2 \, \mathrm{tan}{}^2 \vartheta}{(r^2+l^2)^4}
\Big( r^3 - 3mr^2-3l^2r+ml^2 \Big) \, .
\label{eq:ddr}
\end{equation}
Although we have divided by $\dot{r}$, this equation is valid by 
continuity also if $\dot{r} = 0$. 
By solving simultaneously the equations $\dot{r} =0$ and $\ddot{r}=0$ we
see that there is a circular lightlike geodesic at the intersection of the sphere 
$r=r_{\mathrm{ph}}$ with the cone $\vartheta = \vartheta _{\mathrm{ph}}$ 
where
\begin{equation}
r _{\mathrm{ph}}^3 - 3mr _{\mathrm{ph}}^2-3l^2r _{\mathrm{ph}}+ml^2 
= 0
\label{eq:rph}
\end{equation}
and
\begin{equation}
\mathrm{tan} \, \vartheta _{\mathrm{ph}}= 
\dfrac{r _{\mathrm{ph}}^2+l^2}{2 \, l \, \sqrt{r _{\mathrm{ph}}^2-2mr-l^2}}
\, .
\label{eq:thetaph}
\end{equation}
It is easy to see that the cubic equation (\ref{eq:rph}) has three real solutions
exactly one of which lies in the considered domain, $m + \sqrt{m^2+l^2}
< r_{\mathrm{ph}}$. Applying all possible rotations demonstrates that the sphere
at radius coordinate $r_{\mathrm{ph}}$ is filled with circular lightlike
geodesics. We refer to it as to the \emph{photon sphere}. The existence
of the photon sphere is crucial for determining the socalled \emph{shadow} 
of a NUT black hole, see Grenzebach et al. \cite{arne} where this is 
discussed for a class of spacetimes that contains the NUT spacetime as a
special case.
 
From (\ref{eq:dphitheta}) and (\ref{eq:drtheta}) we get the \emph{orbit
equation} that determines the shape of a lightlike geodesic in the cone 
$\vartheta = \mathrm{constans}$,
\begin{equation}
\dfrac{dr}{d \varphi} = \dfrac{\dot{r}}{\dot{\varphi}} =
\pm 
\, \sqrt{ \dfrac{(r^2+l^2) ^2}{4 \, l^2} \, \mathrm{cos}{}^2 \vartheta 
 - (r^2-2mr-l^2) \, \mathrm{sin}{}^2 \vartheta }
 \, .
\label{eq:orbit}
\end{equation}
From this equation we read that, for $l \neq 0$, there are no lightlike
geodesics contained in the equatorial plane $\vartheta = \pi /2$
because the expression under the square-root must be non-negative.

\section{The deflection angle}\label{sec:defl}
\begin{figure}[H]
    \includegraphics[width=0.7\textwidth]{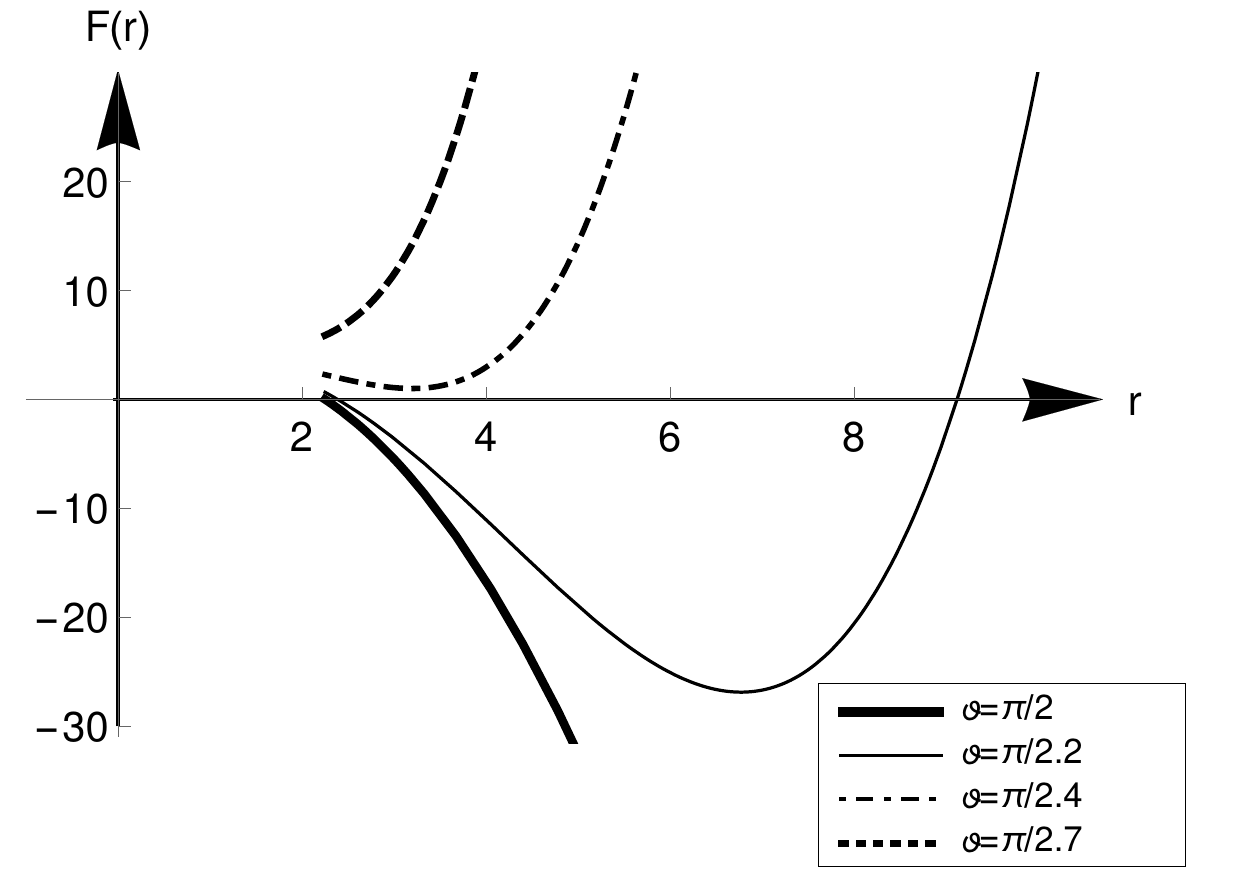}
    \caption{The function $F(r)$ as a function of $\vartheta$ for  
    $l=0.75$, in units with $m=1$ }
    \label{F}
\end{figure}

\begin{figure}[H]
    \includegraphics[width=0.77\textwidth]{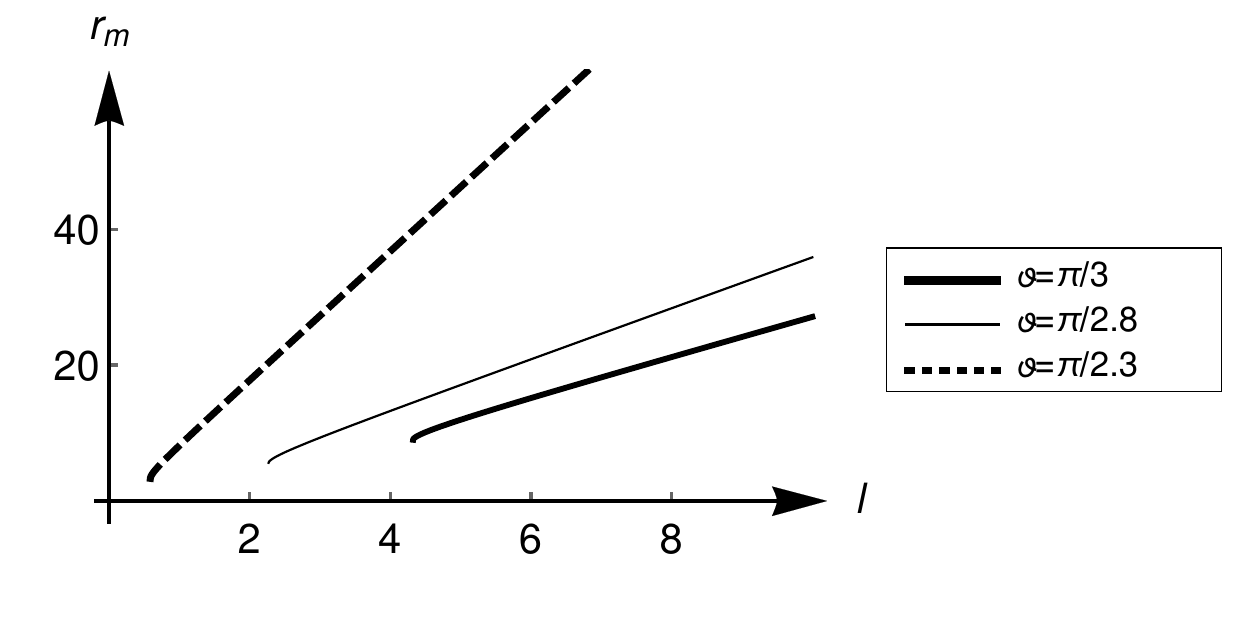}
    \caption{The minimal radius $r_m$ as a function of $l$ for various
    values of $\vartheta$, in units with $m=1$}
\label{rmf1}
\end{figure}

We want to consider, in the cone $\vartheta = \mathrm{constans}$,
a lightlike geodesic that comes in from infinity, goes through a minimum 
radius value $r = r_m$ and then escapes back to infinity. 
Note that such a light ray may make arbitrarily many turns around the
centre. At $r=r_m$, the right-hand side of (\ref{eq:drtheta}) must be 
zero,  
\begin{equation}
\dfrac{r_m^2+l^2}{\sqrt{r_m^2-2 m r_m -l^2}}= 2 \, l \, \tan\vartheta \, ,
\label{eq:rm}
\end{equation}
and the right-hand side of (\ref{eq:ddr}) must be negative. Therefore, 
lightlike geodesics that go through a minimal radius value $r_m$ exist 
for all $r_m$ outside of the photon sphere, $r_m > r_{\mathrm{ph}}$, 
and the corresponding values of $\vartheta$ converge towards 
$\vartheta _{\mathrm{ph}}$ for $r_m \to r_{\mathrm{ph}}$. 
If we define, for fixed $\vartheta$,  the function
\begin{equation}
F(r) := \mathrm{cot} ^2 \vartheta  \, 
( r^2+l^2)^2- 4 l^2 (r^2-2 m r-l^2 ) \, ,
\end{equation}
then $r_m$ is determined by the equation $F(r_m)=0$, where the largest 
positive solution to this fourth-order equation is the relevant one. In Fig.~\ref{F} 
the function $F(r)$ is plotted for a fixed value $l \neq 0$ and various values 
of $\vartheta$. This figure confirms our earlier observation that for $l \neq 0$ there 
are no light rays in the equatorial plane $\vartheta = \pi /2 $.
Fig.~\ref{rmf1} shows a plot of $r_m$ as a function of $l$ for
various values of $\vartheta$. 


With (\ref{eq:rm}) we get from (\ref{eq:orbit}) the  orbit equation in terms 
of $r_m$
\begin{equation}
\dfrac{dr}{d\varphi} 
= 
\pm \, \sqrt{\dfrac{
(r^2+l^2)^2(r_m^2-2 m r_m -l^2)-(r_m^2+l^2)^2(r^2-2 m r-l^2)
}{
(r_m^2+l^2)^2+4l^2(r_m^2-2 m r_m-l^2)
}}
\, .
\label{eq:orbitrm}
\end{equation}
Integration of the orbit equation
(\ref{eq:orbitrm}) over the light ray from its point of closest approach to 
infinity yields
\begin{equation}
\Delta \varphi = \int_{r_m}^{\infty} 
\dfrac{
\sqrt{(r_m^2+l^2)^2+4l^2(r_m^2-2 m r_m-l^2)} \, dr
}{
\sqrt{(r^2+l^2)^2(r_m^2-2 m r_m -l^2)-(r_m^2+l^2)^2(r^2-2 m r-l^2)}
}
\, .
\end{equation}
For defining the deflection angle of a light ray we introduce a new azimuthal 
coordinate 
\begin{equation}
\tilde{\varphi} = \varphi \, \mathrm{sin} \, \vartheta =
\dfrac{(r_m^2+l^2) \, \varphi}{\sqrt{(r_m^2+l^2)^2
+4 l^2 (r_m^2-2 m r_m-l^2)}}
\end{equation}
which is the angle defined around the surface of the cone. On each circle
$r= \mathrm{constans}$ on the cone, $\varphi$ runs from 0 to $2 \pi$
whereas $\tilde{\varphi}$ runs from 0 to $2 \pi \, \mathrm{sin} \, \vartheta$,
i.e., over a smaller interval. This deficit angle can be visualised by cutting the 
cone open and flattening it. 

\begin{figure}[H]
    \includegraphics[width=0.8\textwidth]{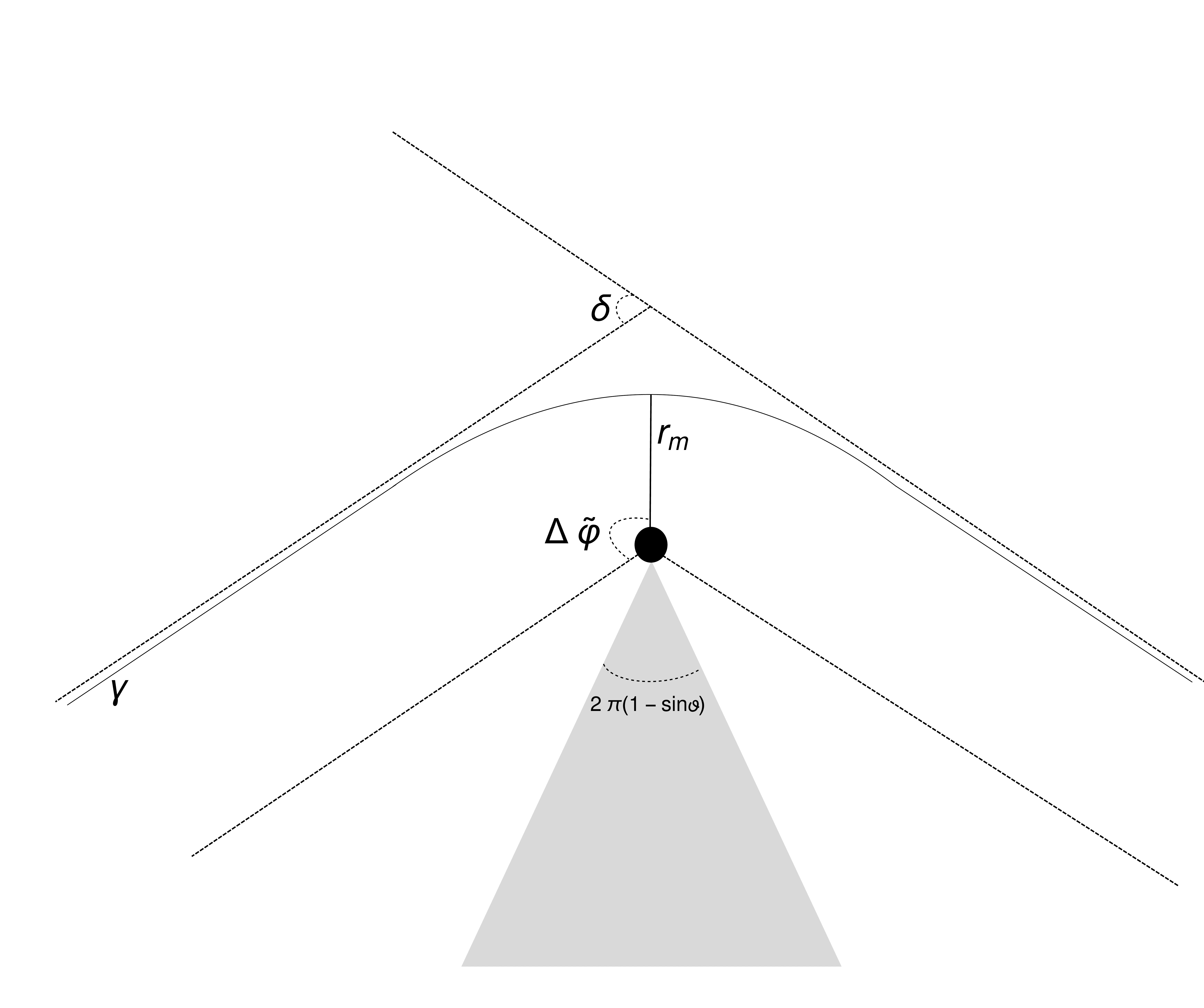}
    \caption{Definition of the deflection angle $\delta$ of a lightlike geodesic
    $\gamma$ in a cone $\vartheta = \mathrm{constans}$}
\label{rm}
\end{figure}

The deflection angle, or bending angle, is defined as the angle under which 
the asymptotes to the light ray intersect in the cut and flattened cone, see 
Fig.~\ref{rm},
\begin{equation}
\delta= 2\Delta \tilde{\varphi} -\pi
\label{def}
\end{equation}
where
\[
\Delta \tilde{\varphi} = \mathrm{sin} \, \vartheta \, \Delta \varphi
\]
\begin{equation}
=
\int_{r_m}^{\infty} 
\dfrac{
(r_m^2+l^2) \, dr
}{
\sqrt{(r^2+l^2)^2(r_m^2-2 m r_m -l^2)-(r_m^2+l^2)^2(r^2-2 m r-l^2)}
}
\, .
\label{eq:Dtphi}
\end{equation}
Here we have used (\ref{eq:rm}).

Our definition of the deflection angle is in agreement with that of 
Nouri-Zonoz and Lynden-Bell \cite{NouriZonoz:1998va}. Note, however,
that in their paper there is a notational inconsistency. In the beginning
they define the NUT metric (with $C=0$) in the same coordinates
as we do: Their $r$ is our $r$ and their $\phi$ is our $\varphi$.
But later, from their eqs. (4) and (5) onwards, their $r$ is our
$\sqrt{r^2+l^2}$ and their $\phi$ is our $\tilde{\varphi}$.
For light rays that make many turns around the centre, the 
deflection angle becomes arbitrarily large, i.e. $\delta \to \infty$
for $r_m \to r_{\mathrm{ph}}$. If, however, $r_m$ is big, $\delta$ is 
small and we get a valid approximation of $\delta$ if we do a Taylor 
expansion of a low order with respect to the dimensionless parameters 
$m/r_m$ and $l/r_m$. We have to go at least up to the 
second order for having a non-vanishing influence of $l$.
Then (\ref{eq:Dtphi}) can be written as 
\[
\Delta \tilde{\varphi} 
=
\int_{r_m}^{\infty} 
\left(
1 \, + \, \dfrac{(r^2+r _m r +r_m^2)}{r (r+r_m)} \, \dfrac{m}{r_m}
\right.
\]
\[
+ \, \dfrac{3 (r^2+r_mr+r_m^2)^2}{2 r^2 (r+m)^2} \, \dfrac{m^2}{r_m^2}
+ \, \dfrac{(3r^2+r_m^2)}{2 r^2} \, \dfrac{l^2}{r_m^2}
\Bigg) \dfrac{r_m \, dr}{r \sqrt{r^2-r_m^2}}  + O(3)
\]
\begin{equation}
=
\dfrac{\pi}{2} + \dfrac{2m}{r_m} - \left( 2 - \dfrac{15 \pi}{8} \right) \dfrac{m^2}{r_m^2}
+ \dfrac{7 \pi}{8} \, \dfrac{l^2}{r_m^2} + O (3) 
\label{eq:Deltatphi}
\end{equation}
where $O(3)$ stands for terms of third or higher order in $m/r_m$
and $l/r_m$. As  a consequence, the bending angle to within this order is
\begin{equation}
\delta =
\dfrac{4m}{r_m} - \left( 4 - \dfrac{15 \pi}{4} \right) \dfrac{m^2}{r_m^2}
+ \dfrac{7 \pi}{4} \, \dfrac{l^2}{r_m^2} + O (3) 
\, .
\label{eq:delta}
\end{equation}
In linear order we have the same result as in the Schwarzschild spacetime
which is clear from the fact that $l$ enters quadratically in all relevant
equations.

\section{Optical metric and Gaussian curvature}\label{sec:optical}
It is known that in an arbitrary spacetime lightlike geodesics satisfy 
a variational principle that can be viewed as a general-relativistic Fermat 
principle, see Temple \cite{Temple1938} for a version restricted to a
local normal neighbourhood, Kovner \cite{Kovner1990} for a formulation
of the general principle and Perlick \cite{Perlick1990a} for a complete 
proof. For static or 
stationary spacetimes, there are simpler and older versions of this 
variational principle. In the static case it was shown already in 
1917 by Weyl \cite{Weyl1917} 
that, as a consequence of Fermat's principle, the spatial paths of lightlike 
geodesics are the geodesics of a Riemannian metric which is now known 
as the \emph{Fermat metric} or the \emph{optical metric}. 
Levi-Civita \cite{Levi1927} generalised Weyl's result to the case of 
a stationary spacetime. Then the spatial paths of lightlike geodesics 
are determined by the combined action of a Riemannian metric and 
a one-form which is sometimes called the \emph{Fermat one-form}; 
for examples we refer to Perlick \cite{Perlick1990b}. Because of 
the influence of the Fermat one-form, the lightlike geodesics do not project to 
geodesics of a Riemannian metric. They rather project to geodesics of a 
Finsler metric. Although, in hindsight, this relation to Finsler geometry 
is fairly obvious, to the
best of our knowledge it was realised only recently, namely by Caponio, 
Javaloyes and Masiello \cite{CaponioEtAl2011} in a paper that was put on 
the arXiv in 2007. It is Levi-Civita's version of Fermat's principle that we will 
apply to the NUT metric now. 
 
According to the foregoing results we may restrict our consideration
to a cone $\vartheta = \mathrm{constans}$. From (\ref{metric}) we read
that along every lightlike \emph{curve} (not necessarily a geodesic) 
the differential of the coordinate time along its spatial projection equals
\begin{equation}
dt= \beta_i dx^i+\sqrt{\bar{g}{}_{ij}dx^i dx^j}
\label{eq:optical}
\end{equation}
where $i,j \in \{r,\varphi\}$. Here $\beta_i$ and $\bar{g}{}_{ij} $ are given by 
\begin{equation}
\beta_i dx^i= 2l (\cos \vartheta+C)d\varphi \; , 
\end{equation}
\begin{equation}
\bar{g}{}_{ij}dx^i dx^j=\dfrac{(r^2+l^2)^2}{(r^2-2mr-l^2)^2}\;dr^2
+\dfrac{(r^2+l^2)^2}{(r^2-2mr-l^2)} \mathrm{sin}{}^2 \vartheta\; d\varphi^2
\, .
\label{eq:opt}
\end{equation}
If one fixes two points in the cone and integrates $dt$ over lightlike curves 
that project to curves in the cone between these two fixed points, 
then the integral over the one-form $\beta _i dx^i$ gives the same value for 
all these curves,
\[
T= \int_{s_1}^{s_2} \beta_i \dfrac{dx^i}{ds} ds
+\int_{s_1}^{s_2} \sqrt{\bar{g}{}_{ij}\frac{dx^i}{ds}\dfrac{ dx^j}{ds}}\; ds
\]
\begin{equation}
= 2l (\cos \vartheta+C) (\varphi(s_2)-\varphi(s_1)) 
+\int_{s_1}^{s_2} \sqrt{\bar{g}{}_{ij}\frac{dx^i}{ds}\dfrac{ dx^j}{ds}}\; ds
\, .
\end{equation}
By Levi-Civita's version of Fermat's principle, for an actual lightlike \emph{geodesic} 
the variation of $T$ vanishes, i.e. 
\begin{equation}
0 = \delta T = 
\delta \int_{s_1}^{s_2} 
\sqrt{\bar{g}{}_{ij}\frac{dx^i}{ds}\dfrac{ dx^j}{ds}}\; ds \, .
\end{equation}
This demonstrates that the spatial path of the lightlike 
geodesic is a geodesic of the Riemannian metric $\bar{g}{}_{ij}$ 
defined in (\ref{eq:opt}). We call it the ``optical metric''
henceforth. Here we have to keep in mind that we have to 
choose the opening angle $\vartheta$ of the cone as determined by 
the impact parameter of the lightlike geodesic according to 
(\ref{eq:J3theta}). As the NUT metric is not static,
but only stationary, it was not to be expected that the
lightlike geodesics project to geodesics of a Riemannian
metric. 

We have already emphasised that calling the surface 
$\vartheta = \mathrm{constans}$ a ``cone'' refers to
the chosen coordinates. The 
optical metric, which has a coordinate-independent meaning, 
makes this surface into a two-dimensional Riemannian manifold.
In order to visualise its intrinsic geometry we try to isometrically
embed  it into Euclidean 3-space. In cylindrical polar coordinates 
($Z$, $R$, $\varphi$), we have to satisfy the condition
\begin{equation}
\bar{g}_{ij}dx^idx^i=dZ^2+dR^2+R^2d \varphi ^2
\end{equation}
with embedding functions $Z(r)$ and $R(r)$. After inserting
(\ref{eq:opt}) and comparing coefficients of $d \varphi ^2$ and 
$dr^2$ we find
\begin{equation}
R(r)^2 =
\dfrac{(r^2+l^2)^2 \mathrm{sin}{}^2 \vartheta}{(r^2-2mr-l^2)} 
\end{equation}
and
\begin{equation}
\Big(\dfrac{d Z(r)}{dr} \Big)^2
= 
\dfrac{
(r^2-2 m r-l^2)(r^2+l^2)^2
- \mathrm{sin} ^2 \vartheta (r^3-3m r^2-3l^2 r+m l^2)^2
}{
(r^2-2m r-l^2)^3
}
\label{eq:emb}
\end{equation}
An isometric embedding into Euclidean 3-space is possible in
that part of the domain $m+\sqrt{m^2+l^2}<r<\infty$ where 
$\big( dZ(r)/dr \big) ^2$ is non-negative. The boundary value, $r=r_b$, of 
the embeddable part is the greatest zero of the sixth-order polynomial
in the numerator on the right-hand of (\ref{eq:emb}). In the 
Schwarzschild limit $l = 0$, which requires $\vartheta =\pi/2$,
we recover the well-known result $r_b = 9 m/4$.

\begin{figure}[H]
    \includegraphics[width=0.8\textwidth]{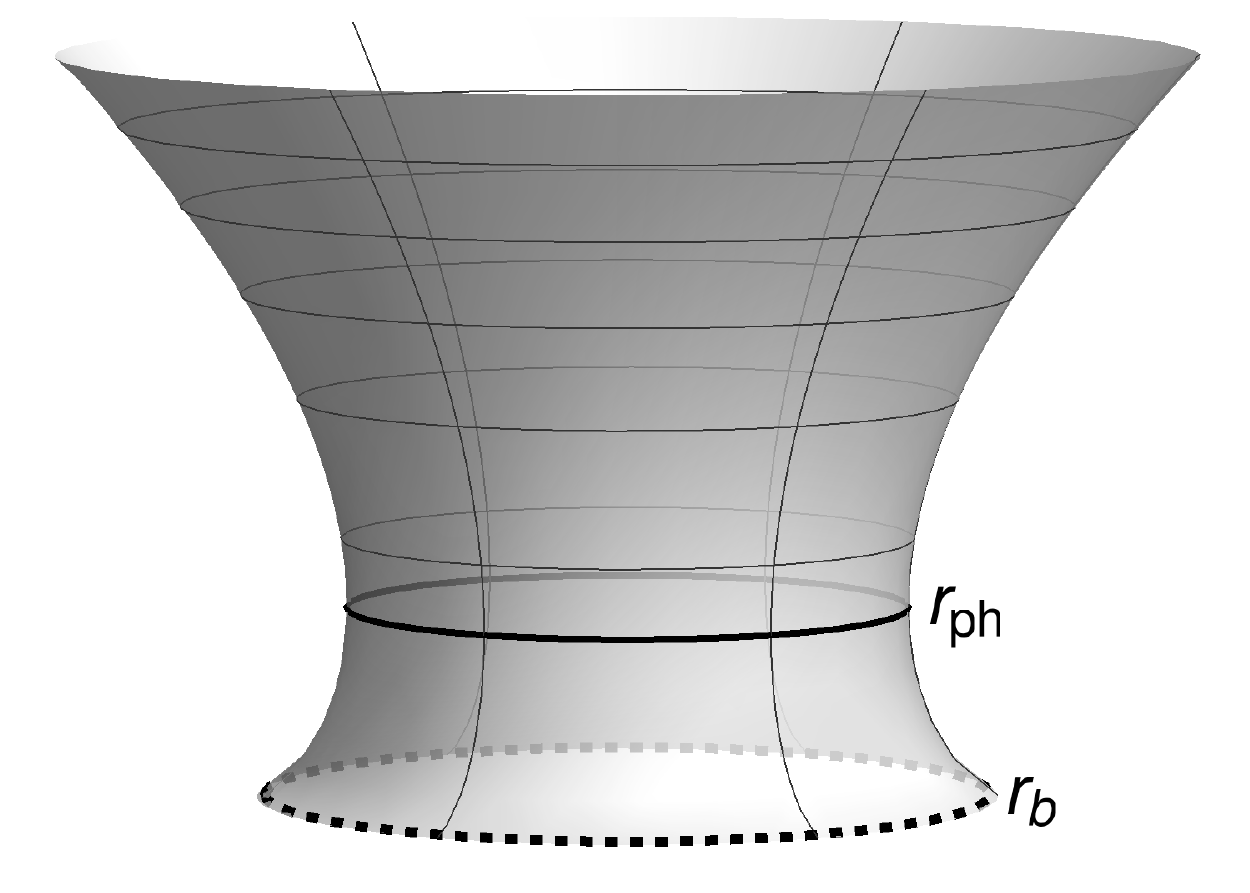}
    \caption{The cone $\vartheta=\pi/3$ of the NUT spacetime with $l=0.5$,   
    embedded as a surface of revolution into Euclidean 3-space, with the photon 
    sphere at $r_{\mathrm{ph}}=3.20 m$ and the boundary of the embeddable 
    part at $r_{b}= 2.38 m$}
\label{em1}
\end{figure}

From Fig.~\ref{em1} we read that the intrinsic geometry of the coordinate 
cone $\vartheta = \mathrm{constans}$ becomes that of a Euclidean cone 
only asymptotically for $r \to \infty$. It develops a ``neck'' at the intersection
with the photon sphere $r= r_{\mathrm{ph}}$, then opens out again before 
the boundary of the embeddable part is reached at $r = r_b$. We also read 
from the picture that at each point the principal curvatures have opposite signs, 
i.e., that the Gaussian curvature is negative which means that geodesics of the 
optical metric must locally diverge. This is true not only for the embeddable 
part but everywhere on the domain $m + \sqrt{m^2+l^2} < r < \infty$ 
on which the optical metric is defined. To demonstrate this we calculate 
the Gaussian curvature of the optical metric analytically.

\begin{figure}[H]
    \includegraphics[width=0.7\textwidth]{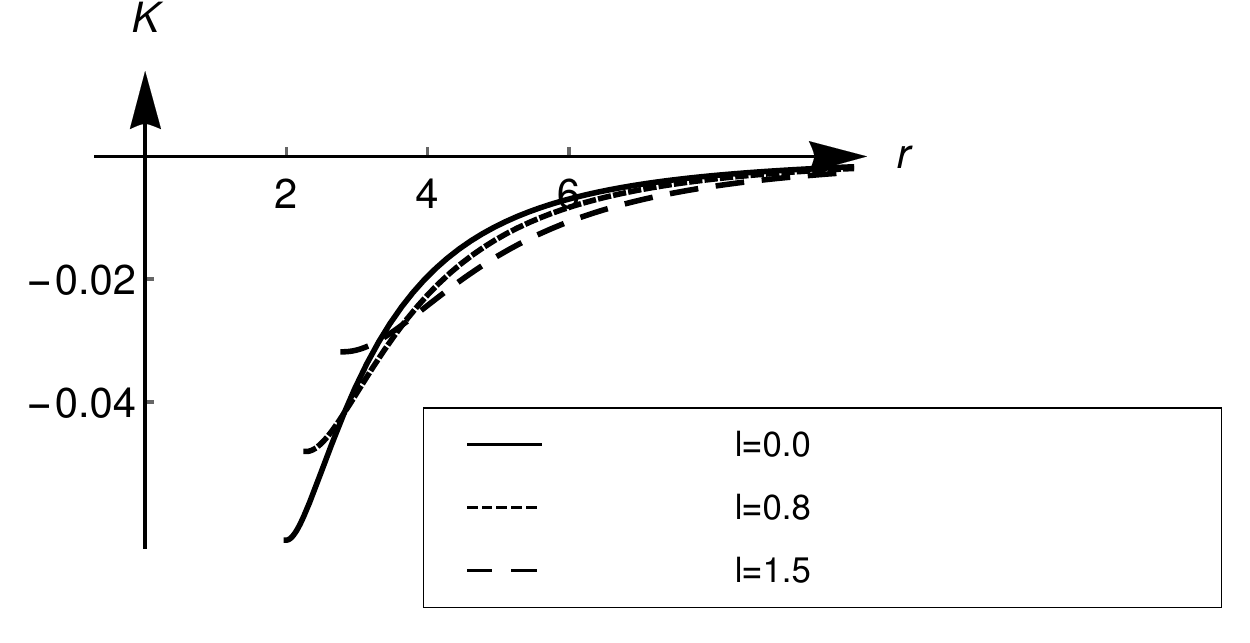}
    \caption{The Gaussian curvature of the optical metric for different NUT parameters,
    plotted in each case for $m + \sqrt{m^2 + l^2} < r$, in units with $m=1$}
    \label{c}
\end{figure}

By definition, the Gaussian curvature $K$ of a two-dimensional
Riemannian metric $\bar{g}{}_{ij}$ satisfies the equation 
\begin{equation}
\bar{R}{}_{ijkl} = K \big( \bar{g}{}_{ik} \bar{g}{}_{jl}- \bar{g}{}_{il} \bar{g}{}_{jk} \big)
\end{equation}
where $\bar{R}{}_{ijkl}$ are the covariant components of the Riemannian curvature tensor.
For a metric of the form (\ref{eq:opt}), the Gaussian curvature can then be calculated as  
\begin{equation}
K = \dfrac{\bar{R}_{r \varphi r\varphi}}{\bar{g}_{rr} \bar{g}{}_{\varphi \varphi}}
\, ,
\end{equation}
which gives, with the special metric coefficients from (\ref{eq:opt}), 
\[
K= - \dfrac{(2 r-3 m ) m \, r^4}{(l^2+r^2)^4}
\]
\begin{equation}
-\dfrac{l^2 r^2 \left(14 m^2-20 m r+7 
r^2\right) + l^4 \left(m^2+10 m r-6 r^2\right) + 3 l^6}{(l^2+r^2)^4} \, .
\label{eq:K}
\end{equation}
For a plot of $K$ as a function of $r$, for different NUT-parameters, see Fig.~\ref{c}.
Eq.~(\ref{eq:K}) confirms that, for any value of $l$, the Gaussian curvature
of the optical metric is indeed negative on the entire domain $m + \sqrt{m^2+l^2}
< r < \infty$. To prove this, substitute $r = m+\sqrt{m^2+l^2}+x$; then
$(l^2+r^2)^4 K$ becomes a fifth-order polynomial in $x$ whose coefficients
are manifestly negative. Also, (\ref{eq:K}) implies that for 
$r \to m + \sqrt{m^2+l^2}$ the Gaussian curvature $K$ approaches a 
strictly negative value and its derivative $dK/dr$ approaches zero. 
Here it is important that we consider a NUT spacetime with $m >0$. The
observation that $K$ is negative is also true for $m =0$ as long as $l \neq 0$.
For $m=0$ and $l =0$, we have Minkowski spacetime and the cones are 
Euclidean cones, i.e, they are locally flat with $K=0$. 

Remarkably, $K$ is independent of $\vartheta$. That is to say, on all cones with
their different opening angles the Gaussian curvature depends on $r$ in 
exactly the same way. A careful look at (\ref{eq:opt}) shows that this is,
actually, not so surprising: If one changes from the coordinates $(r, \varphi )$
to $( \tilde{r} = r , \tilde{\varphi} =  \varphi \, \mathrm{sin} \, \vartheta )$,
the metric coefficients become independent of $\vartheta$. Hence it is clear
that the optical geometries of any two cones with different opening angles
are locally isometric. They are, however, not globally isometric because
the range of the coordinate $\tilde{\varphi}$ depends on $\vartheta$. By 
the same token, their embedding diagrams depend on $\vartheta$ because
in the ambient Euclidean space the azimuthal coordinate is assumed to be
$2 \pi$-periodic. If, on the other hand, we use the representation in a
plane where $(r, \tilde{\varphi})$ are the polar coordinates, as in Fig.~\ref{rm},
then the optical metrics are represented by a metric on this plane that is 
independent of $\vartheta$. We can thus show all the geodesics of the 
optical metrics, i.e., all lightlike geodesics, in one and the same plane; we 
just have to keep in mind that the deficit angle, which is marked in grey 
in Fig.~\ref{rm}, is different for geodesics with different impact parameters.

\section{Gravitational lensing and Gauss-Bonnet theorem}\label{sec:GB}
For the Schwarzschild spacetime, and any other spherically symmetric 
and static spacetime, the Gauss-Bonnet theorem can be used for 
characterising some lensing features. This was demonstrated
in a pioneering paper by Gibbons and Werner \cite{Gibbons:2008rj}
who in particular derived from the Gauss-Bonnet theorem a formula 
for the bending angle  in such spacetimes. Their analysis was based
on the well-known facts that in a spherically symmetric and static
spacetime it suffices to consider lightlike geodesics in the equatorial
plane $\vartheta = \pi /2$ and that the spatial projections of 
lightlike geodesics in this plane are the geodesics of a two-dimensional
Riemannian metric. Werner \cite{Werner:2012rc} demonstrated
that a similar analysis is possible for lightlike geodesics in the 
equatorial plane of the Kerr metric. There are several differences in
comparison to the spherically symmetric and static case. Firstly, 
in the Kerr metric the lightlike geodesics do not project to geodesics
of a Riemannian metric; therefore Werner had to use the idea of an
``osculating Riemannian metric'' for achieving his goal. Secondly,
in the Kerr metric confining oneself to the equatorial plane is, of course, 
a strong restriction of generality. We will now demonstrate that in
the NUT spacetime the Gauss-Bonnet theorem can be applied to 
all lightlike geodesics, without any loss of generality, and that there
is no need to introduce an osculating metric or any other fiducial
quantities. We follow the original idea of Gibbons and Werner as
closely as possible. For background material on the Gauss-Bonnet
theorem the reader may consult any text-book on Riemannian
geometry, e.g. Klingenberg  \cite{Klingenberg}.

\begin{figure}[H]
    \includegraphics[width=0.8\textwidth]{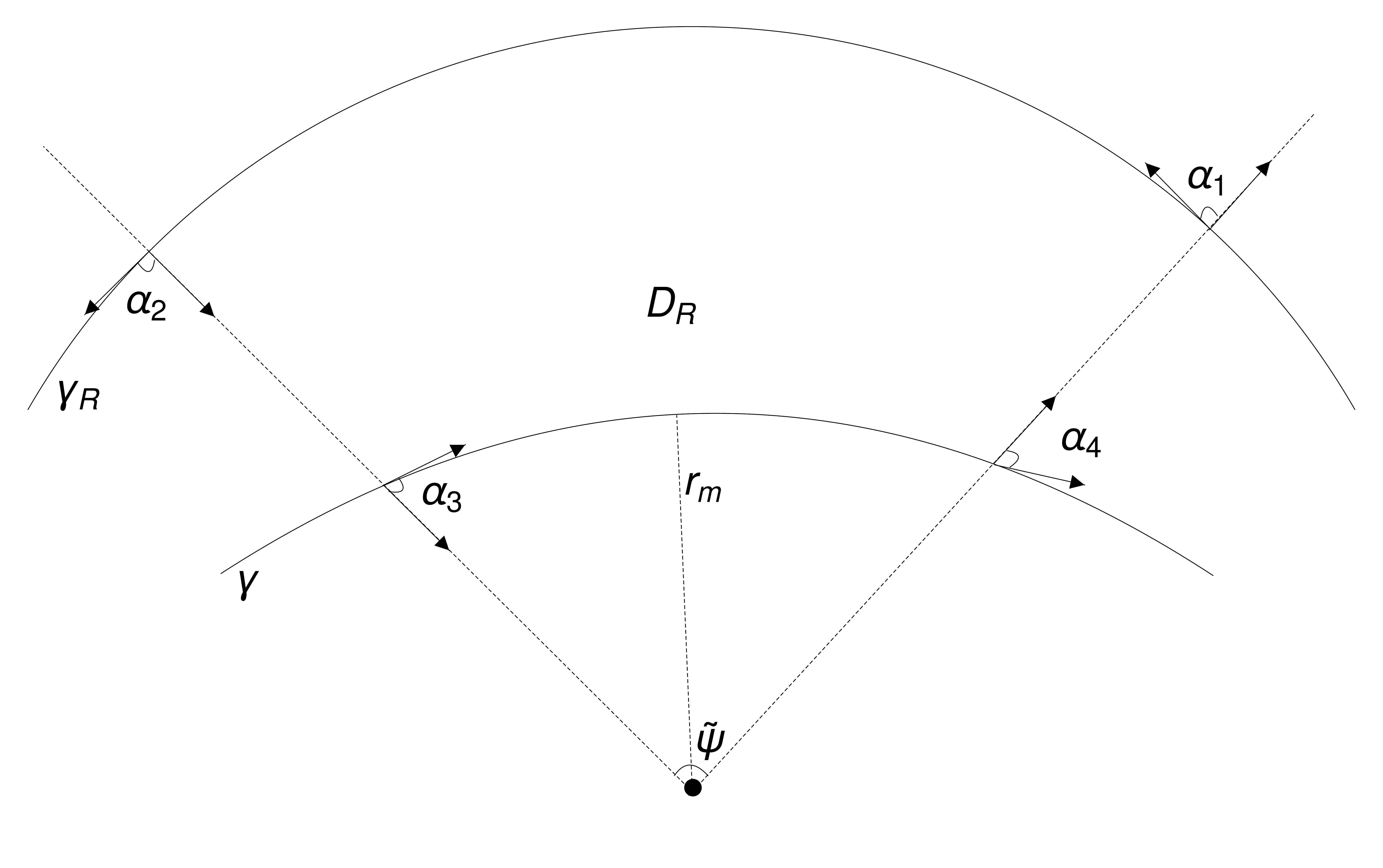}
    \caption{Gravitational lensing and Gauss-Bonnet theorem}
    \label{gauss}
\end{figure}

In its standard text-book version, the Gauss-Bonnet theorem is valid
for a compact domain with boundary in a two-dimensional Riemannian
manifold. For applying this theorem to lightlike geodesics in the NUT
spacetime, we use for the two-dimensional Riemannian manifold
the cone $\vartheta = \mathrm{constans}$ with the optical
metric (\ref{eq:opt}). For the compact domain we use a quadrangle
$D_R$ whose boundary  consists of four smooth parts, see 
Fig.~\ref{gauss}. As in Fig.~\ref{rm}, also in this
picture the cone is shown cut open and flattened, i.e., what is 
shown is a diagram using $(r , \tilde{\varphi})$ as polar coordinates. 

The four smooth parts of the boundary of $D_R$ are constructed in the 
following way. The first part is a finite section of the projection of a lightlike 
geodesic, denoted $\gamma$ in Fig.~\ref{gauss}; we assume that 
$\gamma$ has no self-intersections and that it comes from infinity, goes 
through a minimal radius value $r = r_m$ and then escapes to infinity. 
Along the section of $\gamma$ that is part of the boundary of $D_R$ 
the angle $\tilde{\varphi}$ runs over an interval of length $\tilde{\psi}$.
Another part of the boundary of $D_R$ is a section of a circle $r=R$, 
denoted $\gamma _R$ in the picture. The two remaining parts are 
radial. 

Now the Gauss-Bonnet theorem says that
\begin{equation}
\iint_{D_R} K \, dS
+
\int_{\partial D_R}\kappa \, d s
+ 
\sum_{A=1}^4 \alpha_A=2 \, \pi \, \chi(D_R)
\, .
\label{eq:GB1}
\end{equation}
Here $K$ is the Gaussian curvature and $dS$ is the area element
of the optical metric. $\partial  D_R$ is the boundary of $D_R$
and $\kappa$ is the geodesic curvature of this boundary, to be
integrated over the four smooth parts of $\partial D_R$ where
$s$ denotes arclength with respect to the optical metric. The $\alpha _A$ are 
the four jump angles marked in Fig.~\ref{gauss} and $\chi ( D_R )$
is the Euler characteristic of $D_R$. As $D_R$ is connected and
has no holes, $\chi (D_R) = 1$. Moreover, obviously 
$\alpha _1 = \alpha _2 = \pi /2$. The integral over the boundary 
of $D_R$ reduces to an integral over the circular arc because 
the other three parts are geodesic, hence $\kappa = 0$ there. 
We parametrise the circular arc
by arclength, i.e., we write it in the form $\gamma _R (s)$ 
with $\overline{g} \big( \dot{\gamma _R} , \dot{\gamma _R}) 
= 1$. The geodesic curvature of $\gamma _R$ is then to be 
calculated as 
\begin{equation}
\kappa = 
\big| \overline{g} \big( 
\overline{\nabla}{} _{\dot{\gamma}_R} \dot{\gamma}{}_R, n \big) \big|
\end{equation}
where $\overline{\nabla}$ is the Levi-Civita connection of the optical metric
$\overline{g}$ and $n$ is a unit vector perpendicular to $\dot{\gamma}{}_R$. 
This puts (\ref{eq:GB1}) into 
the following form:
\begin{equation}
\iint_{D_R} K \, dS
+
\int_{\gamma_R}\kappa \, d s
+ 
\alpha _3 + \alpha _4 =  \pi \, .
\label{eq:GB2}
\end{equation}
We now send $R \to \infty$ and $\tilde{\psi} \to 2 \, \Delta \tilde{\varphi}$.
Then obviously $\alpha_3 \to 0$ and $\alpha_4 \to 0$. Moreover, from 
(\ref{eq:opt}) we find after a straight-forward calculation that
\begin{equation}
\int_{\gamma_R}\kappa \, ds \to  2 \, \Delta\tilde{\varphi}
\, .
\end{equation}
Therefore, the limit version of the Gauss-Bonnet theorem reads
\begin{equation}
\iint_{D_{\infty}} K \, dS
+
2 \Delta \tilde{\varphi} 
= \pi \, .
\label{eq:GB3}
\end{equation}
Comparison with \eqref{def} yields
\begin{equation}
\delta=-\iint_{D_{\infty}} K \, dS
\, .
\label{eq:deltaGB0}
\end{equation}
As $K$ is negative, we read from this formula that the deflection angle in the 
NUT spacetime is always positive, i.e., that every light ray is deflected towards
the centre. This is not obvious if the deflection angle is represented with the
help of (\ref{eq:Dtphi})  because one does not know whether the integral on
the right-hand side of this equation is bigger or smaller than $\pi /2$ before 
one has actually calculated it. Moreover, as $K$ is independent of $\vartheta$, 
it is also evident from (\ref{eq:deltaGB0}) that a second light ray has a smaller
deflection angle than the first one if on its entire path it stays farther away from 
the centre. Again, this is not obvious from the representation with 
(\ref{eq:Dtphi}) because the minimal radius $r_m$ occurs not only as the 
lower limit of the integral but also in the integrand. 

When deriving (\ref{eq:deltaGB0}) we have assumed that the light ray has no 
self-in\-ter\-sec\-tion, i.e., we excluded the case that the light ray makes a full
turn, or several full turns, around the centre. If one wants to include such cases, 
one has to consider a region $D_{\infty}$ that is bounded by the parts
of the light ray between infinity and the outermost 
self-intersection point, one has to add $2 n \pi \, \mathrm{sin} \, \vartheta$
on the right-hand side of (\ref{eq:deltaGB0}) where $n$ is the number of full
turns performed by the light ray, and one has to take the jump angle at the
self-intersection point into account. In particular the last point makes the
use of the Gauss-Bonnet theorem to light rays with self-intersections rather
awkward.

To make (\ref{eq:deltaGB0}) more explicit, we write the 
area element of the optical metric as
\begin{equation}
dS=\sqrt{\mathrm{det} \big( \bar{g}_{ij} \big)} \, dr \, d\varphi
=\dfrac{(r^2+l^2)^2 \sin\vartheta \, dr \, d\varphi}{(r^2-2mr-l^2)^{3/2}}
=\dfrac{(r^2+l^2)^2 dr \, d \tilde{\varphi}}{(r^2-2mr-l^2)^{3/2}}
\end{equation}
The bending angle now is 
\begin{equation}
\delta=- 2 \int_0^{\Delta  \tilde{\varphi}} \int_{z( \tilde{\varphi})}^{\infty}
 K \sqrt{\mathrm{det} \big( \bar{g}_{ij}\big) }\;dr \, d \tilde{\varphi}
 \label{eq:deltaGB}
\end{equation}
where  $r=z( \tilde{\varphi})$ is the polar representation of the deflected light ray,
with the angle $\tilde{\varphi}$ as the parameter. 

Werner \cite{Werner:2012rc} has calculated the bending angle in the 
equatorial plane of the Kerr metric to within linear order in the mass 
parameter and the spin parameter. In the NUT metric we have to go 
at least up to second order if we want to have a non-zero contribution 
of the NUT parameter. We have calculated the bending angle already up 
to this order in (\ref{eq:delta}). As  a cross-check, we want to reproduce 
this result with the Gauss-Bonnet theorem.
To that end we need the integrand in (\ref{eq:deltaGB}) up to second order,
\begin{equation}
K \sqrt{\mathrm{det} \big( \bar{g}_{ij}\big) } =
- \dfrac{2m}{r^2}- \dfrac{3m^2}{r^3} - \dfrac{7 l^2}{r^3} + O(3)
\, .
\label{eq:Kdet}
\end{equation}
As this expression has a vanishing zeroth-order term, we need $\Delta \tilde{\varphi}$
and $z ( \tilde{\varphi} )$ only up to first order.  $\Delta \tilde{\varphi}$ was calculated
in (\ref{eq:Deltatphi}),
\begin{equation}
\Delta \tilde{\varphi}
=
\dfrac{\pi}{2} + \dfrac{2m}{r_m} + O (2) \, . 
\label{eq:Deltatphi1}
\end{equation}
For calculating the orbit $r=z ( \tilde{\varphi})$, we integrate on the right-hand side
of (\ref{eq:Deltatphi}) not from $r_m$ but from $z ( \tilde{ \varphi})$ to infinity,
\[
\tilde{\varphi} 
=
\int_{z ( \tilde{\varphi} )} ^{\infty} 
\left(
1 \, + \, \dfrac{(r^2+r _m r +r_m^2)}{r (r+r_m)} \, \dfrac{m}{r_m}
\right) 
\dfrac{r_m \, dr}{r \sqrt{r^2-r_m^2}}  + O(2)
\]
\begin{equation}
= \mathrm{arccos} \Big( \dfrac{r_m}{z ( \tilde{\varphi} )} \Big)
+ \dfrac{\sqrt{z ( \tilde{\varphi} )-r_m}}{\sqrt{z ( \tilde{\varphi} ) + r_m}}
\Bigg( 2 + \dfrac{r_m}{z ( \tilde{\varphi} )} \Bigg) \dfrac{m}{r_m} + O (2)
\end{equation}
where we have assumed that the light ray passes at $\tilde{\varphi} = 0$ 
through the point of closest approach. Solving for $z ( \tilde{\varphi} )$ results
in
\begin{equation}
z ( \tilde{\varphi} ) = \dfrac{r_m}{\mathrm{cos} \, \tilde{\varphi}} 
\left( 1 -
\dfrac{
\big( 2+\mathrm{cos} \, \tilde{\varphi} \big)  \, \mathrm{sin} ^2 \tilde{\varphi}
}{
\big( 1 + \mathrm{cos} \, \tilde{\varphi} \big) \, \mathrm{cos} \, \tilde{\varphi}
}
\, \dfrac{m}{r_m} + O (2) \right)
\label{eq:zphi}
\end{equation}
Inserting (\ref{eq:Kdet}), (\ref{eq:Deltatphi1}) and (\ref{eq:zphi}) into
(\ref{eq:deltaGB}) reproduces, indeed, (\ref{eq:delta}).

We emphasise that the line integral in (\ref{eq:Dtphi}) is easier to 
evaluate than the area integral in (\ref{eq:deltaGB0}). Therefore, if 
the only goal is to actually \emph{calculate} the deflection angle, then
there is no point in using the Gaus-Bonnet theorem.  The merit
of the latter is in the fact that it immediately gives some qualitative
lensing features in terms of geometric quantities.    


\section{Discussion and conclusions}
When Gibbons and Werner \cite{Gibbons:2008rj} applied the
Gauss-Bonnet theorem to the spatial paths of lightlike geodesics 
in static and spherically symmetric spacetimes, many readers 
found this idea attractive because it related the lensing
features in such spacetimes to geometric quantities such as
the Gaussian curvature of a two-dimensional Riemannian metric.  
Naturally, the question arises if a similar result can be found in
spacetimes that are not static and spherically symmetric. Werner
\cite{Werner:2012rc} considered the Kerr metric and he demonstrated 
that, at least for light rays in the equatorial plane, the Gauss-Bonnet 
theorem is applicable. However, in contrast to the static and spherically 
symmetric case, the light rays do not project to geodesics of
a Riemannian metric but rather to geodesics of a Finsler metric 
of Randers type; this made it necessary to introduce a socalled
``osculating metric'' to which the standard Gauss-Bonnet theorem 
could be applied.

Here we have shown that the original method by Gibbons and Werner,
without a restriction to special lightlike geodesics and without the
need of introducing an osculating metric, can be applied to the 
NUT metric. The main difference to the cases considered earlier 
is in the fact that now the spatial projection of each lightlike geodesic
is in a cone, rather than in a plane. Throughout we have allowed for an 
arbitrary value of the Manko-Ruiz parameter $C$, and we have 
verified that $C$ has no influence on the \emph{spatial} paths 
of lightlike geodesics. Our analysis revealed two 
facts that could not have been anticipated before the calculation 
was done: Firstly, we showed that the projection of every lightlike
geodesic is indeed a geodesic of an optical metric that is Riemannian; 
so there is no need of considering Finsler metrics. Secondly, on cones 
with different opening angles the optical metrics turned out to be locally 
isometric; in particular, the Gaussian curvature of the optical metric,
as a function of the radius coordinate, turned out to be independent of 
the opening angle of the cone.  Together with the observation that the 
Gaussian curvature of the optical metric is negative, this allowed us to use  
the Gauss-Bonnet theorem for representing qualitative lensing features 
of the NUT spacetime in a geometric way that is as significant as in the 
Schwarzschild spacetime.

\section*{Acknowledgments}
We gratefully acknowledge support from the DFG within the Research Training Group 1620 ``Models of Gravity''.

\bibliographystyle{spphys}       


\end{document}